\documentclass[aps,prb,twocolumn,superscriptaddress,showpacs]{revtex4}
\usepackage{bm}
\usepackage{amsmath}
\usepackage{amssymb}
\usepackage{graphicx}
\input{epsf}

\begin{document}

\title{Hot electrons in magnetic point contacts as a photon source}
\author{A. M. Kadigrobov}
\affiliation{Department of Physics, University of Gothenburg, SE-412
96 G{\" o}teborg, Sweden} \affiliation{Theoretische Physik III,
Ruhr-Universit\"{a}t Bochum, D-44801 Bochum, Germany}
\author{R. I. Shekhter}
\affiliation{Department of Physics, University of Gothenburg, SE-412
96 G{\" o}teborg, Sweden}
\author{S. I. Kulinich}
\affiliation{Department of Physics, University of Gothenburg, SE-412
96 G{\" o}teborg, Sweden}
\affiliation{B. I. Verkin Institute for Low Temperature Physics and Engineering,
 47 Lenin Avenue, 61103 Kharkov, Ukraine}
\author{M. Jonson}
\affiliation{Department of Physics, University of Gothenburg, SE-412
96 G{\" o}teborg, Sweden} \affiliation{School of Engineering and
Physical Sciences, Heriot-Watt University, Edinburgh EH14 4AS,
Scotland, UK}\affiliation{ Division of Quantum Phases
and Devices, School of Physics, Konkuk University, Seoul 143-701, Korea}
\author{O. P. Balkashin}\affiliation{B. I. Verkin Institute for Low Temperature Physics and Engineering,
 47 Lenin Avenue, 61103 Kharkov, Ukraine}
\author{ V. V. Fisun}\affiliation{B. I. Verkin Institute for Low Temperature Physics and Engineering,
 47 Lenin Avenue, 61103 Kharkov, Ukraine}
\author{ Yu. G. Naidyuk} \affiliation{B. I. Verkin Institute for Low Temperature Physics and Engineering,
 47 Lenin Avenue, 61103 Kharkov, Ukraine}
\author{I. K. Yanson}
\affiliation{B. I. Verkin Institute for Low Temperature Physics and Engineering,
 47 Lenin Avenue, 61103 Kharkov, Ukraine}
\author{S. Andersson} \affiliation{Nanostructure Physics, Royal Institute of Technology, SE-106 91 Stockholm, Sweden}

\author{V. Korenivski}
\affiliation{Nanostructure Physics, Royal Institute of Technology, SE-106 91 Stockholm, Sweden}

\date{\today}

\begin{abstract}
 We propose to use a point contact between a ferromagnetic and a normal metal in the presence of a magnetic field for creating a large inverted spin-population of hot electrons in the contact core. The key point of the proposal is that when these hot electrons relax by flipping their spin, microwave photons are emitted, with a frequency tunable by the applied magnetic field. While point contacts is an established technology their use as a photon source is a new and potentially very useful application. We show that this photon emission process can be detected by means of transport spectroscopy and demonstrate stimulated emission of radiation in the 10-100 GHz range for a model point contact system using a minority-spin ferromagnetic injector. These results can potentially lead to new types of lasers based on spin injection in metals.

\end{abstract}

\pacs{42.55.Ah, 42.55Rz, 73.63.Rt}

\maketitle


\noindent {\bf Introduction} --- Point contact spectroscopy detects relaxation of hot electrons in a small region of a large sample and is a well known method for studying elementary excitations in metals.\cite{Yanson} By using this method the spectral properties of, e.g., phonons and
magnons can be extracted from the bias-voltage dependence of the current flowing through the point contact and, in particular, its component due to the inelastic backscattering of electrons in collisions involving large momentum transfers.

Whether the point contact geometry can be used to make a metal based laser \cite{footnote} is an interesting question that we explore in this Letter. For this to be possible it is necessary to show that photons can be emitted as a continuous flow of hot electrons relax in the contact region and, specifically, that stimulated photon emission leads to a greatly enhanced radiation intensity. This is what we will do in what follows.

Point contact spectroscopy can not be applied to studying of
electron-photon interactions directly, since the momentum of photons is very small. However, by using a point contact between a ferromagnet and a normal metal (or between two ferromagnets) the electron spin comes into play through spin-polarized injection\cite{Rashba,Potok,Chun,Hanbicki,EPL,Zilberman,Wang}.
The spin-split energy bands of the injected electrons can lead to the emmision of photons when the electrons undergo spin-flip relaxation, essentially with no change in the momentum. Since the resistance of spin-up and spin-down channels in magnetic point contacts is different, both emission and absorption
of photons caused by spin-flip inter-channel transitions affects the total current through the contact and, therefore, is detectable by transport spectroscopy.

Below we show that a voltage-biased point contact between a ferromagnet and a normal metal can be used for generating photons with a frequency that can be tuned by means of an external magnetic field through the Zeeman splitting of the spin-up and spin-down energy subbands on the normal metal side of the contact. We show additionally that for realistic magnetic point contacts the spin-flip radiation produced can be detected by means of conventional point contact spectroscopy.

We will first present our model and start with the formalism used for analyzing the transport through the model point contact shown in Fig. 1. The relative weakness of the electron-photon interaction allows us to proceed in two steps. In the first step we calculate the spin populations for the electrons in the contact region, to zeroth order in the electron-photon interaction
strength. In the second step we find the photocurrent in the presence of
radiation as well as the resulting change in the point contact resistance, to first order in the electron-photon interaction.

\noindent {\bf Formalism} --- In order to calculate the electrical current and the spin accumulation one needs to find the  electron distribution function $f_{\sigma \boldsymbol{p}}^{(s)}$ on either side of the contact, i.e., in the ferromagnet ($s$=1) and in the normal metal ($s$=2), for both spin projections $\sigma/2$ ($\sigma=\pm 1$) as a function of position $\boldsymbol{r}$
and electron momentum $\boldsymbol{p}$. In each of the metals these functions satisfy  the Boltzmann equation
\begin{eqnarray}
\boldsymbol{v}_\sigma^{(s)}\frac{\partial f_{\sigma \boldsymbol{p}}^{(s)}}{\partial \boldsymbol{r}}-e\frac{\partial \Phi^{(s)}}{\partial \boldsymbol{r}}\frac{\partial f_{\sigma \boldsymbol{p}}^{(s)}}{\partial \boldsymbol{p}} +\frac{f_{\sigma \boldsymbol{p}}^{(s)} -\langle f_{\sigma \boldsymbol{p}}^{(s)}\rangle}{\tau_\sigma^{(s)}} \nonumber \\ =\sigma w_{ph}^{(s)}\{f^{(s)}_{\uparrow \boldsymbol{p}},f^{(s)}_{\downarrow \boldsymbol{p}}\};\;s=1,2\,.
 \label{Boltzmanngeneral}
\end{eqnarray}
Here $\boldsymbol{v}^{(s)}_\sigma$ is the electron velocity in each metal, which is given
 by a momentum derivative of the electron energy as
 $\boldsymbol{v}^{(s)}_\sigma=\partial  E_\sigma^{(s)} (\boldsymbol{p})/\partial\boldsymbol{p}$,
 where $E_\sigma^{(1)} (\boldsymbol{p})=\varepsilon^{(1)} (\boldsymbol{p})-\sigma J_1$ and
$E_\sigma^{(2)} (\boldsymbol{p})=\varepsilon^{(2)} (\boldsymbol{p})+\sigma \mu_B H $ both
contain a spin independent kinetic energy term $\varepsilon^{(s)} (\boldsymbol{p})$. For convenience of notation,  $\sigma =+1$ and $\sigma =-1$ here and below correspond to the directions of the electron magnetic moment parallel and antiparallel to the magnetization direction in the ferromagnet ($s=1$), respectively (that is the electron magnetic moment projections in the ferromagnet are $m_\sigma^{(1)} =\mu_B \sigma/2$). The spin
dependence of the energy in the normal metal ($s$=2) is due to a static external magnetic field
$H$ and the associated Zeeman energy  gap $2\mu_B H $ (here and below we assume the
electron g-factor $g=2$), while the much stronger spin-dependence in the ferromagnetic
metal ($s$=1) is due to the exchange energy $J_1$ (the Zeeman energy can be neglected
here). Furthermore $\tau_\sigma^{(s)} = l_0/|\boldsymbol{v}_\sigma^{(s)}| $ is the elastic
relaxation time and $l_0$ is the elastic mean free path of the electrons in the
diffusive transport regime considered in this paper, $\Phi^{(s)}$ is the electric potential in the respective metal
and the notation $\langle\ldots\rangle$ implies an average of the bracketed quantity over
the  Fermi surface.
 \begin{figure}
  \centerline{\includegraphics[width=0.8\columnwidth]{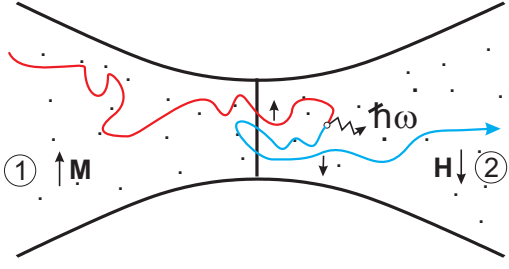}}
  \caption{Diffusive point contact under irradiation (not shown) in the presence of a static magnetic field  $H$. A voltage bias $V$ injects a spin polarized current from a ferromagnetic metal
  (1)  with magnetic moment $M$ into a normal metal (2). An electron with its magnetic moment up (spin-down)
  is shown to move along a diffusive trajectory from  metal 1 to metal 2 (red line) where it resonantly interacts with the irradiation field, which results in a spin flip and the emission of a photon. The electron continues along its diffusive path with the magnetic moment down    (blue line) thereby changing the spin-dependent contact resistance.}
   \label{PC}
  \end{figure}

 The amplitude of the electromagnetic field irradiating the point contact is assumed to be large enough to allow the
 electron-photon interaction to be treated semi-classically, so that the collision integral in
 Eq. (\ref{Boltzmanngeneral}) can be written as
 \begin{eqnarray}
w_{ph}^{(s)}\{f^{(s)}_{\uparrow \boldsymbol{p}},f^{(s)}_{\downarrow \boldsymbol{p}}\}=
 \frac{2 \pi}{\hbar}|\mu_B h_{ac}|^2 \left[f^{(s)}_{\downarrow\boldsymbol{p}-\boldsymbol{q}}-f^{(s)}_{\uparrow \boldsymbol{p}}\right]\times
 \nonumber\\
 \delta \Bigl(E^{(s)}_\uparrow(\boldsymbol{p})-E^{(s)}_\downarrow(\boldsymbol{p}-\boldsymbol{q})-\hbar\omega \Bigr)
 \label{el-phot collision integral}
\end{eqnarray}
Here $h_{ac}$ is the magnetic amplitude of  the electromagnetic wave of frequency
$\omega$ and momentum $|\boldsymbol{q}|=\hbar \omega/c$ that irradiates the point contact;
$c$ is the velocity of light.

To facilitate explicit calculations we will consider a simplified contact geometry,
 approximating  the point contact by a cylindrical channel of length $L$ and diameter
 $d$, with  $L\gg d \gg l_0$.
The boundary conditions for the electron distribution functions  $f^{(s)}_{\sigma \boldsymbol{p}}$
at the interface  between metals 1 and 2  can be written in the form
\begin{eqnarray}
f^{(1)}_{\sigma \boldsymbol{p}}=(1-D_{\sigma})f^{(1)}_{\sigma \boldsymbol{p}_R}+D_{\sigma}f^{(2)}_{\sigma \boldsymbol{p}_T}\nonumber \\
f^{(2)}_{\sigma \boldsymbol{p}}=D_{\sigma}f^{(1)}_{\sigma \boldsymbol{p}_T}+(1-D_{\sigma})f^{(2)}_{\sigma \boldsymbol{p}_R}
 \label{boundary}
\end{eqnarray}
where $D_\sigma = D_{\sigma}(\boldsymbol{ p},\boldsymbol{ p}_T)$ is the spin-dependent
transparency of the interface; $\boldsymbol{ p} = (\boldsymbol{ p}_\parallel,\; p_z)$ and
$\boldsymbol{ p}_R=(\boldsymbol{ p}_\parallel,\;- p_z)$ (here $\boldsymbol{ p}_\parallel=(p_x,p_y)$)
are the momenta of the incident and reflected electrons, respectively; the momentum of the transmitted
electron  $\boldsymbol{ p}_T$ is determined by the condition of the energy conservation
$E^{1,2}_\sigma (\boldsymbol{ p})=E^{2,1}_\sigma (\boldsymbol{ p}_T)$.

Away from the contact region the current spreads over a large volume so that its density
decreases and the electron system is essentially in equilibrium at distances $\boldsymbol{| r}| \gg d$.
In our simplified geometry we will therefore use the additional boundary conditions
$f^{(1,2)}_{\sigma \boldsymbol{p}}\left(z=\pm L/2\right) = n_F(E_\sigma^{(1,2)} (\boldsymbol{p}))$
where $n_F$ is the Fermi distribution function and the $z$-axis is directed along the point contact.

We will now solve the electron-photon scattering problem formulated above in the weak
scattering limit characterized by $d/\l_{ph} \ll 1$, where $l_{ph}$ is the electron-photon
scattering length. This allows us to solve the Boltzmann equations (\ref{Boltzmanngeneral})
by perturbation theory, where $w^{(s)}_{ph}$,  $f_{\sigma \boldsymbol{p}}^{(s)}$, and
$\Phi^{(s)}$ are expanded in powers of the small parameter $d/\l_{ph}$. We will first solve
the problem to zeroth order in $d/\l_{ph}$, which allows us to find the density of hot electrons
with an inverse spin population in the contact region, and then solve for the photocurrent to
first (linear) order in $d/\l_{ph}$.

\begin{figure}
  \centerline{\includegraphics[width=0.8\columnwidth]{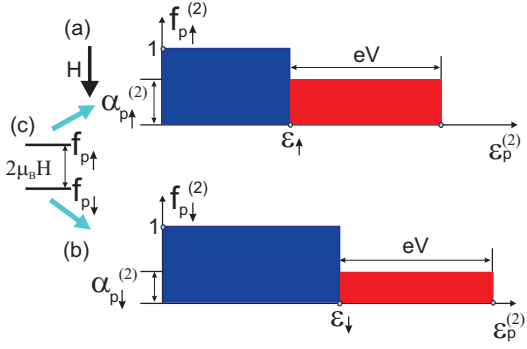}}
  \caption{Zero-temperature energy distributions for (a)  magnetic moment-up (spin-down), $f_{\boldsymbol{p}\uparrow}$,  and (b)  magnetic moment-down   (spin-up) electrons,
  $f_{\boldsymbol{p}\downarrow}$, at point $\boldsymbol{r}$ on the normal-metal side of the point contact. The inset (c)
  shows the Zeeman energy splitting and the direction of the magnetic field $H$. All states are occupied up to
$\varepsilon_{\uparrow}= \varepsilon_f -eV/2 -\mu_B H$ and $\varepsilon_{\downarrow}= \varepsilon_f -eV/2 +\mu_B H$,
respectively (blue rectangles), but in the intervals $(\varepsilon_{\uparrow},\varepsilon_{\uparrow}+eV)$ and
$(\varepsilon_{\downarrow},\varepsilon_{\downarrow}+eV)$ the states are only partly occupied
(red rectangles) and to an extent that is determined by the probabilities $\alpha_{\uparrow \boldsymbol{p}}(\boldsymbol{r})$ and
$\alpha_{\downarrow \boldsymbol{p}}(\boldsymbol{r})$
for ``hot" electrons in the ferromagnet to reach $\boldsymbol{r}$.
Clearly, the difference between the densities of spin-down and spin-up electrons, $ n_\uparrow (\boldsymbol{r})-n_\downarrow (\boldsymbol{r})
\propto [(\alpha_\uparrow^{(2)} -\alpha_\downarrow^{(2)}) eV -2  \mu_B H ]$, depends on the bias voltage $V$.
It follows that the spin population can be inverted, so that $n_\uparrow (\boldsymbol{r})>n_\downarrow (\boldsymbol{r})$, for large enough $V$
if  $\alpha_\uparrow^{(2)} > \alpha_\downarrow^{(2)}$.
}
   \label{Pumping}
  \end{figure}

{\bf Spin accumulation} --- In order to solve the kinetic equations (\ref{Boltzmanngeneral})
to zeroth order in $d/\l_{ph} $ we generalize the procedure developed in
Refs. \cite{Robert1,Robert2,Robert3} to allow for spin dependent electron dynamics.
To zeroth order, the  distribution functions $f^{(s)}_{\sigma \boldsymbol{p}}$ can be written as
\begin{eqnarray}
f^{(s)}_{\sigma \boldsymbol{p}}= \alpha_{\sigma \boldsymbol{p}}^{(s)}n_F\left(E_\sigma^{(s)} (\boldsymbol{p})
+
e\phi_0(\boldsymbol{r}) -eV/2\right)
\nonumber\\
+(1-\alpha_{\sigma \boldsymbol{p}}^{(s)})n_F\left(E_\sigma^{(s)} (\boldsymbol{p})+e\phi_0(\boldsymbol{r}) +eV/2\right)\,,
\label{falpha}
\end{eqnarray}
where $\alpha_{\sigma \boldsymbol{p}}^{(s)}(\boldsymbol{r})$ is the probability that an
electron emanating from far inside the ferromagnet ($z=-\infty$) diffuses elastically to reach
point $\boldsymbol{r}$ in metal $s$ with momentum $\boldsymbol{p}$; the concrete form of
the electrical potential  $\phi_0(\boldsymbol{r})$ inside the point contact is not important in
the limit $eV \ll \varepsilon_F$. The distribution functions $f^{(2)}_{\sigma \boldsymbol{p}}$
are sketched in Fig.~\ref{Pumping}.

To linear order in the parameter $l_0/d\ll 1$, it follows from Eqs. (\ref{Boltzmanngeneral}) and
(\ref{boundary}) that this probability can be expressed as
$\alpha^{(s)}_{\sigma \boldsymbol{p}}= \langle\alpha_{\sigma \boldsymbol{p}}^{(s)}\rangle - l_0 (v_z/|\boldsymbol{v}|) d \langle\alpha_{\sigma \boldsymbol{p}}^{(s)}\rangle/d z$. The isotropic part of $\alpha^{(s)}_{\sigma \boldsymbol{p}}$ satisfies the diffusion equation
\begin{equation}
 \frac{d^2}{d z^2} \langle\alpha_{\sigma \boldsymbol{p}}^{(s)}\rangle =0,
\label{alphadiffusion}
\end{equation}
 with the boundary conditions   $\langle\alpha_{\sigma \boldsymbol{p}}^{(1)}(z=-L/2)\rangle=1$ and $\langle\alpha_{\sigma \boldsymbol{p}}^{(2)}(z=L/2)\rangle=0$;
in the vicinity of the F/N interface  the effective boundary conditions are\cite{Robert2}
\begin{eqnarray}
\langle\alpha_{\sigma \boldsymbol{p}}^{(2)}\rangle-\langle\alpha_{\sigma \boldsymbol{p}}^{(1)}\rangle&=&\frac{l_0}{\langle D_\sigma\rangle}\frac{ d \langle\alpha_{\sigma \boldsymbol{p}}^{(1)}\rangle}{d z};
\nonumber\\
 \frac{d\langle\alpha_{\sigma \boldsymbol{p}}^{(1)}\rangle}{d z}&=&\frac{d \langle\alpha_{\sigma \boldsymbol{p}}^{(2)}\rangle}{d z};\hspace{0.7cm}
 \label{effectiveboundary}
\end{eqnarray}
if the transparency of the interface is assumed to be small, $\langle D_\sigma\rangle\ll 1$.
Solving the diffusion equation (\ref{alphadiffusion}) with these  boundary conditions one finds
\begin{equation}
\langle\alpha_\sigma^{(1)}\rangle=1-\beta_\sigma^{(1)}(1+\frac{2 z}{ L}); \;\;\;\langle\alpha_\sigma^{(2)}\rangle=\beta_\sigma^{(2)}(1-\frac{2z}{ L})\,,
\label{alpha}
\end{equation}
where
\begin{equation}
\beta_\sigma^{(s)}=\frac{\kappa^{(s)}_\sigma}{1+\kappa^{(1)}_\sigma+\kappa^{(2)}_\sigma}; \hspace{0.3cm}
\kappa^{(s)}_\sigma=\langle D_\sigma\rangle\frac{L}{2 l_0}.
\label{beta}
\end{equation}

If  electrons are injected from the ferromagnet (1) into the normal metal (2) (i.e., if $eV > ~0$) the number of ``hot" electrons with spin up $\uparrow$  and down $\downarrow$ that accumulate in the effective volume $\Omega_{PC}^{(2)}\sim d^3$ of normal metal in the point contact (PC) is
\begin{eqnarray}\label{deltansigma}
&&\delta n_{\sigma}=\int_{\Omega_{PC}^{(2)}} d^3\boldsymbol r
\int\frac{d^3\boldsymbol p}{(2\pi \hbar)^3}\times\nonumber\\
&&\left[f_{\boldsymbol p\sigma}^{(2)}(\boldsymbol r)-n_F
\left(E_\sigma^{(2)} (\boldsymbol p)+e\phi_0(\boldsymbol r)+
eV/2\right) \right].
\nonumber
\end{eqnarray}
Using  Eqs. (\ref{falpha}), (\ref{alpha}), and (\ref{beta}) this expression can be evaluated to give
\begin{equation}
\delta n_\sigma =\frac{\beta_{\sigma}^{(2)}}{2}\Bigl(n_0 \Omega_{PC}^{(2)} \Bigr)\frac{eV}{ \varepsilon_F}\,,
\label{deltansigma}
\end{equation}
where  $n_0$ is the conduction electron density in the normal metal.

From the result (\ref{deltansigma}) we conclude that the total number of hot electrons injected into the normal-metal side of the contact is
\begin{equation}
\delta n=\gamma_{tr}\Bigl(n_0 \Omega_{PC}^{(2)} \Bigr)\frac{eV}{ \varepsilon_F};\;\;  \gamma_{tr} =\frac{\beta_\uparrow^{(2)}+\beta_\downarrow^{(2)}}{2 }
\label{deltantotal}
\end{equation}
and the induced magnetic moment corresponding to the net spin density accumulated in the same region is
\begin{eqnarray}
\delta M =\mu_B S \delta n=\mu_B\beta_{tr}\Bigl(n_0 \Omega_{PC}^{(2)} \Bigr)\frac{eV}{2 \varepsilon_F}\,.
\label{deltaM}
\end{eqnarray}
Here  $S$, the effective spin of an injected  electron, is
\begin{equation}
S=\frac{1}{2}\frac{\delta n_\uparrow - \delta n_\downarrow}{ \delta n} = \frac{1}{2} \frac{\beta_\uparrow^{(2)}-\beta_\downarrow^{(2)}}{\beta_\uparrow^{(2)}+\beta_\downarrow^{(2)}}\equiv \frac{1}{2} \beta_{tr}
\label{spin}
\end{equation}
and $\beta_{tr}$ is a measure of the spin polarization of the hot electrons injected into the contact region.
If the size $d$ of the contact is a few times 10 nm and $\beta_{tr}\sim 0.3$, which corresponds to a nearly ballistic
point contact with $d \sim l_0$ and a spin polarization of 30\% at the F/N interface, the injected number of hot
electrons and the induced magnetic moment in the contact region is $\delta n \sim 10^6 \,eV/\varepsilon_F$ and
$\delta M \sim 10^6 \mu_B\, eV/\varepsilon_F$, respectively.

{\bf Photocurrent} --- The previous calculations can readily be extended to find the photocurrent flowing through the
point contact under irradiation. Since the electron-photon interaction hardly affects the electron momentum at all,
the main cause of the photocurrent is the photon-induced electron spin-flip transitions in conjunction with the
spin-dependence of the contact resistance. The spin flips change the electron spin densities in the contact and
the spin-dependent contact resistance is connected with the different  densities of states for the two spin projections.

In order to find the photocurrent we first solve the Boltzmann equation (\ref{Boltzmanngeneral}) for the
photon-induced change $f_{\sigma \boldsymbol{p},1}^{(s)}(\boldsymbol{r})$ in the electron distribution function.
We do so to lowest (linear) order in the small parameter $d/\l_{ph}$
and with  the boundary conditions $f_{\sigma \boldsymbol{p},1}^{(1,2)}(z=\mp L/2)=0$.
The matching conditions at the F/N interface are given by Eq. (\ref{boundary}) with the change $f_{\sigma \boldsymbol{p}}^{(s)} \rightarrow   f_{\sigma \boldsymbol{p},1}^{(s)} $.
Using these solutions one finds
the photocurrent  as
\begin{eqnarray}
\label{photocurrentgeneral1}
I_{ph}&=&
    e\sum_{s=1,2}\int_{\Omega_{PC}^{(s)}} d\boldsymbol{r} \int \frac{d \boldsymbol{p}}{(2\pi \hbar)^3} \times\\
    && \Bigl(\alpha^{(s)}_{\uparrow, -\boldsymbol{p}} (\boldsymbol{r})- \alpha^{(s)}_{\downarrow, -\boldsymbol{p}} (\boldsymbol{r}) \Bigr)w^{(s)}_{ph}\{f^{(s)}_{\uparrow  \boldsymbol{p}0},f^{(s)}_{\downarrow  \boldsymbol{p}0} \}.
   \nonumber
\end{eqnarray}
Using Eqs. (\ref{photocurrentgeneral1}), (\ref{alpha}) and (\ref{beta}) one obtains the total current $I(V)$ in a diffusive
point contact under irradiation as
\begin{eqnarray}
I(V)&=&\frac{V}{R}+\theta\left(1-\frac{|\hbar \omega-2\mu_B H|}{\hbar \omega}\frac{c}{v_F}\right)j_{ph}(V);
\nonumber \\
j_{ph}(V)&=&\frac{\Delta R}{R^2}\left( V-V^*\right)
 \label{totalcurrent}
\end{eqnarray}
Here $R$ is the ``dark" contact resistance due mainly to the impurities, while the relative change of the
point contact resistance caused by the irradiation is
\begin{eqnarray}
\frac{\Delta R}{R}=\frac{ (2\pi \bar{\beta}_{tr})^2}{6}\frac{c}{v_F}\frac{| \mu_B h_{ac}|^2}{\varepsilon_F \hbar \omega}(n_0 \Omega_{PC}^{(2)})\Bigl(\frac{2e^2}{h}R\Bigr)\,,
\label{resistancechange}
\end{eqnarray}
where $\bar{\beta}_{tr}=\beta_{tr}\gamma_{tr}$ and $ eV^*= (3/4)\hbar \omega/\bar{\beta}_{tr}$. As one sees from Eq. (\ref{totalcurrent}) the dependence of the photocurrent on the magnetic field has a peak corresponding to the resonant interaction of the electron spin and the electromagnetic field.

A comparison between Eq. (\ref{photocurrentgeneral1}) and the rate equation for photons generated by  electronic spin-flip transitions induced by the electromagnetic field (see \cite{EPL}),
 \begin{equation}
  \frac{d n_{ph}^{(s)}}{d t}= -\int w_{ph}^{(s)}\{f^{(s)}_{\uparrow \boldsymbol{p}},f^{(s)}_{\downarrow \boldsymbol{p}}\}\frac{d^3\boldsymbol{p}}{(2\pi\hbar)^3}\,,
 \label{rate equation}
 \end{equation}
where $n_{ph}$ is the photon density,
shows that the photocurrent may be re-written in the form
\begin{eqnarray}
I_{ph}=
    -e\sum_{s=1,2}\int_{\Omega_s} d\boldsymbol{r} \Bigl(\langle\alpha^{(s)}_{\uparrow,\boldsymbol{p}}\rangle - \langle\alpha^{(s)}_{\downarrow, \boldsymbol{p}}\rangle\Bigr)
 \frac{d n_{ph}^{(s)}}{d t}\,,
\label{photocurrentemissionrate}
\end{eqnarray}
which makes it clear that its magnitude depends on the net rate of photon absorption/emission
in combination with the spin dependence of the effective transparency of the point contact. From  Eq. (\ref{totalcurrent}) one notes that the microwave-induced current changes sign at $V=V^*$, i.e. when the rate of photon emission by ``hot" electrons begins to exceed the rate of photon absorption.

The close association between the electron transport and  photon radiation processes allows us to express the photocurrent in terms of the power
of emission and absorption of photons by electrons in the point contact. Using Eqs. (\ref{el-phot collision integral}), (\ref{falpha}), and (\ref{alpha}) one finds that the
net emitted power due to resonant ($\hbar \omega =2 \mu_B H$) absorption and
emission of photons  in the irradiated point contact, defined as $P(V) = \hbar \omega \int d\boldsymbol{r} d n_{ph}/d t $, can be expressed as
\begin{equation}
P(V)=P_{0}\left(-1 +\frac{3}{2}\frac{V}{V^*}\right)\,.
\label{powerV}
\end{equation}
Here
\begin{equation}
P_{0}= \frac{\pi}{2} \frac{ c}{ v_F}\Bigl(n_0\Omega_{PC}^{(2)} \Bigr)\frac{| \mu_B h_{ac}|^2}{\varepsilon_F}\omega
\label{power0}
\end{equation}
is the absorbed power due to photon absorption, while the second term in Eq.(\ref{powerV}) is the emitted power due to photon
emission from the point contact.

Comparing Eq. (\ref{totalcurrent}) and Eq. (\ref{power0}) one finds that
\begin{equation}
j_{ph}(V)=\frac{3}{4}\frac{V-V^*}{V^{*2}}P_{0}\,,
\label{current-power}
\end{equation}
which makes it possible to find the power $P_{0}$ absorbed from the electromagnetic field by measuring
$dj_{ph}(V)/dV$ (see Eq.~(\ref{totalcurrent})) after first having determined $V^*$ from the condition
$j_{ph}(V^*)=0$. Furthermore, the net emitted power $P(V)$ can be determined by measuring
$j_{ph}(V)$ with the help of  Eq.~(\ref{current-power}) and Eq.~(\ref{powerV}).

{\bf So far,} we have shown theoretically that an inverted spin population is accumulated in a voltage biased point contact  between a ferromagnet (F) and a normal metal (N). For a contact of linear dimension $d \sim 10$~nm, biased by  a voltage $V$, and with a spin polarization of 30\% at the F/N interface ($\beta_{tr}\sim 0.3$) we find that the corresponding magnetic moment injected into the contact region is  $\delta M \sim 10^6\mu_B  eV/\varepsilon_F$.
We have furthermore shown that if the point contact is irradiated by an electromagnetic field, photon-induced electron spin-flip scattering gives rise to a peak in the relative change of the point contact  resistance, $\Delta R/R$ as a function of the irradiation frequency. The peak appears when the frequency is resonant with the Zeeman splitting in the normal-metal spectrum of conduction electrons, which for an external magnetic field of 1 T occurs at 30~GHz. The net power, $P(V)$, generated by the stimulated emission of photons in the electron spin-flip relaxation process can be determined by measuring the photon current $j_{ph}(V)$ defined in Eq.~(\ref{totalcurrent}). In the experiment discussed below with typically ~10~mW, 10-100~GHz microwaves irradiating the point contact produces a magnetic amplitude ($h_{ac}$) of the electromagnetic field inside the point contact of approximately 30~mT. For such a field we find that $\Delta R/R \sim $ 0.01--0.10\% and that $P(V)$ is given by Eq.~(\ref{powerV}) with $P(0) \sim$ 1--10~pW; $P(0)$ being the power absorbed from the electromagnetic field due to photon absorption in the contact region. These estimates show that an experimental implementation of the proposed spin-laser effect in magnetic point contacts is feasible, and is demonstrated below. Two comments are in order. The neglect of spin-flip scattering in the normal metal due to magnetic impurities or spin-orbit interaction is well justified since the point contact size of ~10 nm is one to two orders of magnitude smaller than the spin-diffusion length in a typical normal metal such as Cu. Another possible imperfection is spin perturbations that can occur in nano-constrictions, especially in external fields applied opposite to the magnetization in the ferromagnetic electrode (needed for spin-population inversion). One would need a very hard ferromagnet unaffected even at the interface by a reversing field field of ~1 T, such as transition metal-rare earth alloy. There is another, rather interesting solution employing a spin-minority injector, where possible spin perturbations are actually fully suppressed by the high Zeeman field. This latter configuration is illustrated experimentally below.

{\bf Stimulated photon emission in FeCr/Cu point contacts} --- In this section we provide experimental evidence for the effect described above. The system is a point contact of tip-surface type, between a ferromagnetic film and a nonmagnetic tip made of Cu. The ferromagnetic material chosen for this experiment is a known minority-carrier Fe70Cr30 alloy \cite{Fert}, in which the majority of the conduction electrons have their magnetic moments opposite to the local magnetization and therefore spins parallel to the local magnetization (conduction electrons with magnetic moments parallel to the local magnetization are a minority). This inverse spin polarization is approximately 30\% for the chosen alloy composition. The use of a minority-spin injector offers a rather special configuration for creating a strong Zeeman splitting, which is desirable in detecting the laser effect discussed above. An external magnetic field applied parallel to the magnetization of the minority injector automatically is anti-parallel to the net injected magnetic moment in Cu. This produces the desired spin-population inversion, with the high-energy level more populated by the injected polarized electrons then the low-energy level. Since the field is parallel to the injector's magnetization, its magnitude can be increased arbitrarily high and the spin-population inversion condition would only improve. The field of a few Tesla yields a Zeeman splitting of the order of 100 GHz, well outside the 1-10 GHz range, typical for spin-torque dynamics. (We note also that the bias current of order of 10 $\mu$ A used below is two orders of magnitude lower than that typically required for inducing spin-torque effects -- see, e.g., our recent results \cite{Balkashin} for details). This allows to effectively separate the spin-photonic effects discussed herein from the spin-torque effects in the system. Furthermore, a high field in excess of 2 T leads to an essentially perfect magnetic alignment in the ferromagnet, including the interfacial spins in the nano-constriction, which greatly simplifies the interpretation of the experiment. This minority-injector configuration of stimulated spin-flip photo-emission is illustrated in Fig. 3a.

 \begin{figure}
  \centerline{\includegraphics[width=0.8\columnwidth]{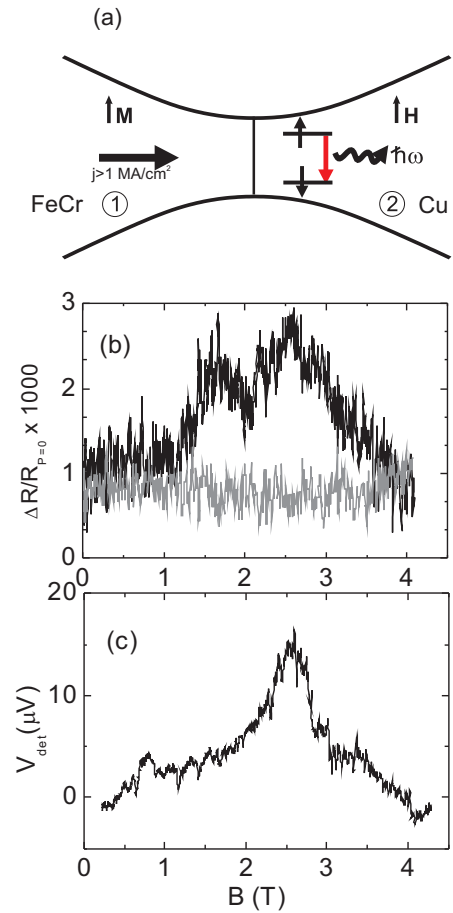}}
  \caption{
  (a) Schematic illustration of spin-flip photon emission in F/N point contact, with the ferromagnetic electrode of spin-minority type. The injected electrons are spin-split by the external field applied parallel to the magnetization of the injector, and the spin-up and spin-down levels are inversely populated. (b) Resistance of a point contact un-irradiated (grey curve) and irradiated (black curve),
  $\Delta R/R_0$ : Cu100/Fe70Cr3050/Cu3-Cu, $f$= 64 GHz, $P$ = 10 mW, $T$=4.2 K, $V_{bias}$=-0.16 mV, $R_0=R(P=0,V\rightarrow 0)$=10.6 $\Omega$ corresponding to the point contact diameter of $\sim $10 nm. (c) Detector voltage locked-in to the chopping frequency (2 kHz) of the irradiating microwave field (64 GHz), directly proportional to the change in the point contact resistance under irradiation: $T$=4.2 K, $V_{bias}$=-0.16 mV,  $R_0=20 \Omega $ corresponding to $d_{PC} \sim$ 5 nm. (b) and (c) are for two different point contacts.
  }
   \label{Exp}
  \end{figure}

The experimental arrangement in terms of producing point contacts and microwave irradiating them were discussed in detail in our recent publications \cite{Yanson,Balkashin}. Here we concentrate on the region in the phase space of the system, in terms of the bias current and irradiation frequency, where previously reported effects of spin-torque dynamics are absent. The irradiation frequency is 64 GHz, corresponding to the Zeeman field of approximately 2.3 T for a free electron (appropriate for Cu). The resistance with the microwave power off is essentially flat within the noise floor of the measurement, in the entire field range of 4 T. This background is subtracted from the resistance measured with the microwave power on. The difference is then normalized and shown in Fig. 3b as a function of field. The resistance becomes bell-shaped under irradiation, centered around 2.5 T, corresponding well to the expected Zeeman splitting at 64 GHz. The magnitude of the measured $\Delta R/R$ is of the order of 0.1\%, which also agrees well with the above theoretical predictions for this point contact geometry. In order to improve the signal to noise ratio, the microwave power was chopped at 2 kHz and the resistance measured using a lock-in amplifier referenced to the chopping  frequency. The lock-in signal is then directly proportional to the difference in resistance with and without the irradiation. The resulting detector voltage measured for a FeCr/Cu point contact (different from that in b) is shown in Fig. 3c. A pronounced peak in the vicinity of the expected Zeeman splitting is evidence for a relaxation process stimulated by the microwave field. This process has a resonant character in terms of its field-frequency condition, coinciding with that for stimulated photon emission by spin-flip relaxation.

{\bf In conclusion,} we have shown that a suitably implemented spin injection can be used to achieve a tunable photon emission by a metal. This effect has a great potential for new types of spin-based lasers, which are expected to have extremely high optical gain compared for example to semiconductor lasers \cite{EPL}.

{\bf Acknowledgements.}
Financial support from the European Commission (FP7-ICT-FET proj no 225955 STELE), the Swedish VR, and the Korean WCU program funded by MEST/NFR (R31-2008-000-10057-0) is gratefully acknowledged.

\end{document}